# Attitudes toward Open Access, Open Peer Review, and Altmetrics among Contributors to Spanish Scholarly Journals

**(Preprint)**

Francisco Segado-Boj, Juan Martín-Quevedo, Juan José Prieto-Gutiérrez

# Attitudes toward Open Access, Open Peer Review, and Altmetrics among Contributors to Spanish Scholarly Journals

**Abstract**

This paper aims to gain a better understanding of the perspectives of contributors to Spanish academic journals regarding open access, open peer review, and altmetrics. It also explores how age, gender, professional experience, career history, and perception and use of social media influence authors' opinions toward these developments in scholarly publishing. A sample of contributors (n = 1254) to Spanish academic journals was invited to participate in a survey about the aforementioned topics. The response rate was 24 per cent (n = 295). Contributors to Spanish scholarly journals hold a favourable opinion of open access but were more cautious about open peer review and altmetrics. Younger and female scholars were more reluctant to accept open peer review practices. A positive attitude toward social networks did not necessarily translate into enthusiasm for emerging trends in scholarly publishing. Despite this, ResearchGate users were more aware of altmetrics.

**Introduction**

Technological innovation has changed how academic journals are edited and published, how manuscripts submitted to them are reviewed, and how the impact of their published contents is measured.[1] Technological innovation has also raised expectations of universal access to knowledge.[2] The phenomenon having the greatest impact on scholarly publishing today is the open access (OA) movement.[3] To be considered an OA publication, a journal must allow authors to self-archive their articles in free-access institutional or specialist repositories or make their work available on their own websites (referred to as green open access); alternatively, journals can make the content freely available on the platforms that host their content, sometimes requiring authors to pay an article processing charge(referred to as gold open access).[4]

These same technological innovations have given rise to calls for more transparency and broader public participation in science.[5] Internet-based communication systems have made an open review process a feasible option.[6] The term *open peer review* refers to a range of practices that include allowing authors and reviewers to be aware of each other's identities, publishing articles and reviewer commentaries side by side, and encouraging broader community involvement before and/or after publication.[7]

Furthermore, new methods for measuring the reach and uptake of scientific output that consider more than journal and article impact have been developed. Altmetrics, for example, measure the impact of scholarly research from fresh perspectives such as the number of times an article has been mentioned in blogs or shared via social media.[8]

However promising these innovations may seem to their adopters and exponents, academic authors' perceptions of them determine the degree of their willingness to engage with them.[9] Recent studies on attitudes toward OA in Spain and other countries indicate that researchers publishing today generally feel positive about OA,[10] and they perceive OA's greatest advantage to be the opportunity it offers to disseminate their work to wider audiences.[11] Obversely, the misgivings most frequently expressed by researchers concern the quality of journals employing OA systems and the credibility of OA publications,[12] as well as the transfer of publishing costs to the author.[13]

We undertook a survey study to measure the awareness of and attitudes toward these new publishing practices held by academic authors whose research has been published in Spanish scholarly journals. Ours is the first large-scale survey study, to our knowledge, to analyze not only academic authors' opinions regarding OA but their perceptions of altmetrics and open peer review as well. Although another study recently published surveyed Spanish researchers' attitudes toward OA publishing, that study focused principally on the experimental sciences.[14] We believe that the research we report here, which addresses additional facets of scholarly communication and covers a broader range of disciplines, constitutes a valuable addition to the existing literature.

One of the virtues of this study is its scope, which is not limited to one disciplinary domain of knowledge,[15] a particular group of universities,[16] a single institution,[17] or authors with publications in journals owned by the same publishing group.[18] We consider a broader array of variables that may influence authors' opinions than have studies focused more narrowly on issues such as age,[19] academic rank,[20] academic experience,[21] and discipline.[22]

Communication research indicates that individuals' media choices and habits influence the type of content they consume and the manner in which they consume it.[23] We undertook our study on the premise that scholars' use of and attention to media, specifically online media and services, may relate to their perceptions of emerging trends in scholarly communication. We wanted to test for possible associations between scholars' perceptions of the aforementioned innovations in academic publishing and the ways in which these individuals use internet services in general and social-networking platforms in particular.

**Research Questions**

The main objective of this study was to gain a better understanding of the perspectives of contributors to Spanish academic journals regarding changes taking place in scholarly communication. Another objective of this study was to determine the degree to which a particular set of variables (age, gender, professional experience, and perceptions and use of social-networking platforms) influence authors' opinions of changes afoot in the way scholarly articles are now being reviewed and published. With these objectives in mind, we formulated the following research questions to guide our study:

1. What do authors who contribute to Spanish scholarly journals think about OA?

2. What do these authors think about open peer review?

3. What do they think about altmetrics?

4. How do the variables, listed parenthetically above, affect their opinions of OA, open peer review, and altmetrics.

**Methodology**

The survey we employed in this study was divided into sections. The first section asked socio-demographic questions about the respondent's gender, age, academic experience, country of residence, and main area of knowledge per Scopus subject-area classifications. Later, those subject-areas were grouped into four main categories, mimicking the structure of the Spanish national agency for evaluation of Higher Education (ANECA): arts and humanities, social sciences, engineering and sciences. In a second section devoted to Web 2.0 services (that is, social media websites and scholarly-networking platforms), respondents were asked to indicate whether they maintained personal blogs or accounts in Twitter, Facebook, LinkedIn, Academia.edu (abbreviated AE hereinafter), ResearchGate (abbreviated RG hereinafter), or Mendeley. The next section employed an intensity scale,[24] that is, a Likert scale anchored at 0

(low end; negative) and 6 (high end; positive) to measure the respondent's engagement on Facebook, Twitter, and AE and RG (these last two were grouped together for the purposes of this part of the survey). Respondents were then asked whether they were familiar with the emerging trends in scholarly publishing described in our introduction. Those who responded yes were asked to indicate their level of agreement with a series of statements, extracted from a qualitative study of opinions and attitudes held by editors of Spanish scholarly journals on the same topics.[25]

Author email addresses were gathered from a screening sample of 1279 articles published in 2015 and 2016 by fifteen scholarly journals sampled following previous studies[26], all of them affiliated or edited by an Spanish institution as registered in the 'country' field in Scopus database. Language was not considered a criterion, and Spanish-based journals which publish papers in English were included in the sample. The sample comprised journals indexed at least in one the following subject categories: Architecture, Building and Construction, Civil & Structural Engineering, Communication, Education, Geography, Planning and Development, History, History & Philosophy of Science, Language & Linguistics, Law, Library & Information Science, Mechanics of Materials, Medicine, Political Science & International Relations, Psychology, Sociology & Political Science & Sports Science. Most of the selected journals were related to Social Sciences disciplines, which may explain, as discussed below the high presence of scholars from these fields among survey respondents.

As some authors published more than one article in the considered time period, their email appeared more than once. Those duplicates e-mails were identified and merged into one only record, so that authors received only one invitation to complete the survey, the final list of authors to whom a survey was sent totalled 1254. Questionnaires were sent to individuals on this list by email on 12 February 2017. To ensure that an adequate number of recipients responded, a reminder message urging participation was sent to everyone on this list on 20 February.

Recipients were given until February 28 to complete the online questionnaire, created and submitted via 'Forms' app in Google Drive. Survey was originally submitted in Spanish, so that the statements that appear in Tables 2-4, have been translated into English from its original version.

Of the 1254 authors contacted, 295 (24 per cent) participated (see Table 1 for a socio-demographic description of the participants). This figure is comparable to the percentages reported by the authors of several recent similar studies, which were 26 per cent,[27] 27 per cent,[28] and 23 per cent.[29] To facilitate comparison between categories, responses to the statements with a graduated scale—that is, the intensity, or Likert, items on the survey—were divided into two groups. Answers 0–2 on the scale were classified as negative and answers 3–5 as positive. (Note that not all participants answered every question, so counts within the categories presented in tables do not always total 295; moreover, percentages were rounded to whole numbers.)

Table 1. Socio-demographic overview of survey participants

| Participant Characteristic | Number | Percentage |
| --- | --- | --- |
| Gender | | |
|    Male | 162 | 55 |
|    Female | 129 | 44 |
| Age in years | | |
|    30 or younger | 30 | 10 |
|    31–40 | 71 | 24 |
|    41–50 | 99 | 33 |
|    51–60 | 74 | 25 |
|    61 or older | 21 | 7 |

| Career experience in years | | |
|---|---|---|
| 5 or fewer | 49 | 17 |
| 6–15 | 101 | 34 |
| 16–25 | 92 | 31 |
| 26 or more | 52 | 18 |
| Country of residence | | |
| Spain | 246 | 83 |
| Mexico | 9 | 3 |
| Brazil | 8 | 3 |
| Chile | 8 | 3 |
| Colombia | 5 | 2 |
| Argentina | 4 | 1 |
| Portugal | 3 | 1 |
| Italy | 2 | 1 |
| Other* | 8 (1 per) | 3 |
| Disciplinary domain | | |
| Social sciences | 187 | |
| Arts and humanities | 47 | |
| Engineering | 41 | |
| Sciences | 16 | |
| Web service used | | |
| RG | 181 | 61 |
| Facebook | 168 | 57 |
| AE | 133 | 45 |
| Twitter | 123 | 42 |
| LinkedIn | 118 | 40 |
| Mendeley | 64 | 22 |
| Personal blog | 44 | 15 |

\* The other countries (eight in all) in which only one participant reported residing were Australia, Belgium, Ecuador, France, Morocco, the Netherlands, the United Kingdom, and the United States.

**Results**

Although the overwhelming majority of survey participants (92 per cent) knew about OA, far fewer were familiar with open peer review (65 per cent) and even fewer had a working understanding of altmetrics (41 per cent).

*Open Access*

The statements on which respondents across the board expressed the highest degree of agreement concerned the positive aspects of OA. As for its downsides, respondents' agreement with statements fell below 50 per cent or even 20 per cent. In Table 2, a total of 272 (92 per cent) respondents agreed with statement "I know what open access is" (see column [1]). Respondents who agreed with the other six statements concerning OA—columns [2] to [7]—represent a subset of the respondents summed in column [1]. So the percentages in columns [2] to [7] were calculated with the number in column [1] as the denominator. This is true for the values in all rows of Table 2.

Table 2. Respondent agreement with statements concerning open access

| | [1] | [2] | [3] | [4] | [5] | [6] | [7] |
|---|---|---|---|---|---|---|---|
| Participant Variable | no. (%) | no. (%) | no. (%) | no. (%) | no. (%) | no. (%) | no. (%) |
| All | 272 (92) | 258 (95) | 241 (89) | 178 (65) | 47 (17) | 246 (90) | 119 (44) |
| Social Sciences | 170 (91) | 163 (96) | 148 (87) | 111 (65) | 23 (14) | 155 (91) | 66 (39) |
| Gender | | | | | | | |
|     Male | 148 (91) | 142 (96) | 133 (90) | 96 (65) | 24 (16) | 132 (89) | 59 (40) |
|     Female | 119 (92) | 111 (93) | 103 (87) | 78 (66) | 23 (19) | 109 (92) | 59 (50) |
| Age (years) | | | | | | | |
|     ≤30 | 25 (83) | 24 (96) | 24 (96) | 17 (68) | 2 (8) | 23 (92) | 9 (36) |
|     31–40 | 66 (100) | 66 (100) | 60 (91) | 46 (70) | 14 (21) | 64 (97) | 29 (44) |
|     41–50 | 91 (92) | 86 (95) | 80 (88) | 60 (66) | 14 (15) | 84 (92) | 34 (37) |
|     51–60 | 70 (95) | 62 (89) | 58 (83) | 43 (61) | 15 (21) | 56 (80) | 36 (51) |
|     ≥61 | 19 (90) | 19 (100) | 18 (95) | 11 (58) | 2 (11) | 18 (95) | 10 (53) |
| Experience (years) | | | | | | | |
|     ≤5 | 42 (86) | 42 (100) | 42 (100) | 32 (76) | 6 (14) | 42 (100) | 17 (40) |
|     6–15 | 95 (94) | 89 (94) | 78 (82) | 60 (63) | 17 (18) | 84 (88) | 40 (42) |
|     16–25 | 86 (93) | 81 (94) | 79 (92) | 55 (64) | 14 (16) | 78 (91) | 42 (49) |
|     ≥26 | 48 (92) | 45 (94) | 41 (85) | 30 (63) | 10 (21) | 41 (85) | 20 (42) |
| Use Twitter? | | | | | | | |
|     Yes | 114 (93) | 110 (96) | 104 (91) | 79 (69) | 16 (14) | 106 (93) | 46 (40) |
|     No | 158 (91) | 148 (94) | 137 (87) | 99 (63) | 31 (20) | 140 (89) | 73 (46) |
| Use Facebook? | | | | | | | |
|     Yes | 155 (92) | 148 (94) | 140 (90) | 79 (69) | 26 (17) | 106 (93) | 46 (40) |
|     No | 117 (91) | 110 (94) | 101 (86) | 73 (62) | 21 (18) | 99 (85) | 52 (44) |
| Use LinkedIn? | | | | | | | |
|     Yes | 115 (97) | 112 (97) | 105 (91) | 78 (68) | 19 (17) | 106 (92) | 55 (48) |
|     No | 157 (88) | 145 (92) | 135 (86) | 99 (63) | 27 (17) | 139 (89) | 64 (41) |
| Have personal blog? | | | | | | | |
|     Yes | 43 (98) | 43 (100) | 40 (93) | 31 (72) | 7 (16) | 40 (93) | 19 (44) |
|     No | 229 (91) | 214 (93) | 200 (87) | 146 (64) | 39 (17) | 205 (90) | 100 (44) |
| Use AE? | | | | | | | |
|     Yes | 130 (98) | 125 (96) | 120 (92) | 89 (68) | 27 (21) | 122 (94) | 56 (43) |
|     No | 142 (88) | 133 (94) | 121 (85) | 89 (63) | 20 (14) | 124 (87) | 63 (44) |
| Use RG? | | | | | | | |
|     Yes | 176 (97) | 169 (96) | 162 (92) | 115 (65) | 34 (19) | 160 (91) | 84 (48) |

[1] I know what OA is.
[2] OA facilitates access to scientific and scholarly knowledge.
[3] OA promotes scholarly debate.
[4] For publishers, OA makes less profit than pay walls and subscription models.
[5] Unlike pay walls and subscriptions, OA undermines the scholarly value of articles.
[6] OA increases the visibility of articles and their probability of citation.
[7] In OA, authors must pay to publish.

|  | | | | | | | |
|---|---|---|---|---|---|---|---|
| No | 96 (84) | 89 (93) | 79 (82) | 63 (66) | 13 (14) | 86 (90) | 35 (36) |
| Use neither RG nor AE | 59 (79) | 54 (92) | 47 (80) | 40 (68) | 8 (14) | 52 (88) | 22 (37) |
| Use Mendeley? | | | | | | | |
| Yes | 62 (97) | 60 (97) | 59 (95) | 44 (71) | 12 (19) | 58 (94) | 36 (58) |
| No | 120 (91) | 197 (94) | 181 (86) | 133 (63) | 34 (16) | 187 (89) | 83 (40) |
| Pos. attitude to Facebook | 49 (92) | 45 (92) | 42 (86) | 30 (61) | 4 (8) | 45 (92) | 20 (41) |
| Neg. attitude to Facebook | 223 (92) | 210 (94) | 196 (88) | 145 (65) | 42 (19) | 198 (89) | 97 (44) |
| Pos. attitude to Twitter | 60 (94) | 55 (92) | 52 (87) | 38 (63) | 19 (32) | 50 (83) | 29 (48) |
| Neg. attitude to Twitter | 212 (91) | 203 (96) | 189 (89) | 140 (66) | 28 (13) | 196 (92) | 90 (42) |
| Pos. attitude to RG & AE | 136 (99) | 134 (99) | 131 (96) | 90 (66) | 26 (19) | 126 (93) | 62 (46) |
| Neg. attitude to RG & AE | 136 (86) | 131 (96) | 128 (94) | 88 (65) | 24 (18) | 123 (90) | 60 (44) |
| Standard deviation | 59 (4) | 56 (3) | 52 (5) | 38 (4) | 10 (4) | 53 (4) | 25 (5) |

Respondents across the board also tended to agree with the statement that OA offers the public greater access to scientific findings: 89 per cent of respondents aged 51–60 endorsed this assertion (column [2]). Early-career academics tended to agree with the statement that OA facilitates access to scientific knowledge to a somewhat greater degree than other respondents. Age and academic experience tended to influence the attitudes expressed by survey participants. Younger respondents and those at an early stage of their careers agreed that OA increases the visibility of articles and their possibility of being cited. Older authors and those with more academic experience were a bit more sceptical about the potential of OA to increase the visibility of scholarly output and the number of citations (column [6]).

Mendeley users were more apt to agree that OA makes authors pay for publishing their articles than other categories of study participant. The percentage of older respondents that agreed with this assertion was also above the average for the sample as a whole. Somewhat curiously, respondents aged 41–50 were relatively less inclined to support this assumption. Participants who did not use RG or AE were least apt to believe that OA made authors to pay to publish (column [7]).

Respondents aged 31–40 (70 per cent) as well as those with less than five years of career experience (76 per cent) agreed more strongly with the statement 'From a publisher's perspective, OA is less profitable than pay wall and subscription models' than other study participants (column [4]). Respondents who maintained personal blogs or used Mendeley also expressed higher than average levels of support for this opinion (72 and 71 per cent, respectively).

### *Open Peer Review*

More than half of the participants in this study felt that a deeply rooted cultural resistance to criticism was likely to impede broad acceptance of open peer review in the countries they lived and worked in. They nevertheless perceived the advantages and disadvantages of the open peer review to be roughly equal. The statement in this section of survey with which they disagreed the most was the assertion that open peer review evaluations were likely to be tainted by personal biases and grudges. In Table 3, a total of 192 (65 per cent) respondents agreed with statement 'I know what open peer review is' (see column [1]). Respondents who agreed with the other six statements concerning open peer review—columns [2] to [7]—represent a subset of the respondents summed in column [1]. As with Table 2, the percentages in columns [2] to [7] were calculated with the number in column [1] as the denominator. This is true for the values in all rows of Table 3.

Table 3. Respondent agreement with statements concerning open peer review

| [1] I know what open peer review is. |
| [2] Open peer review is more likely to be tainted by personal biases and grudges. |
| [3] Reviewers are naturally reluctant to criticize their peers openly. |
| [4] Open peer review can improve the quality of research findings. |
| [5] Such a system would be hard to implement in my country given cultural resistance to criticism. |
| [6] Authors may manipulate the system by recruiting colleagues expected to give positive evaluation. |
| [7] Open peer review enhances the transparency of the evaluation process. |

| Participant Variable | [1] | [2] | [3] | [4] | [5] | [6] | [7] |
|---|---|---|---|---|---|---|---|

|  | no. (%) | no. (%) | no. (%) | no. (%) | no. (%) | no. (%) | no. (%) |
|---|---|---|---|---|---|---|---|
| All | 192 (65) | 91 (47) | 116 (60) | 115 (60) | 124 (65) | 112 (58) | 116 (60) |
| Social Sciences | 124 (67) | 53 (43) | 74 (60) | 76 (61) | 80 (65) | 73 (59) | 80 (65) |
| Gender | | | | | | | |
|     Male | 105 (65) | 47 (45) | 65 (62) | 61 (58) | 65 (62) | 55 (52) | 59 (56) |
|     Female | 84 (65) | 42 (50) | 48 (57) | 52 (62) | 57 (68) | 56 (67) | 55 (65) |
| Age (years) | | | | | | | |
|     ≤30 | 17 (57) | 7 (41) | 9 (53) | 11 (65) | 7 (41) | 10 (59) | 9 (53) |
|     31–40 | 46 (65) | 19 (41) | 26 (57) | 30 (65) | 32 (70) | 26 (57) | 29 (63) |
|     41–50 | 64 (65) | 33 (52) | 38 (59) | 36 (56) | 44 (69) | 45 (70) | 41 (64) |
|     51–60 | 54 (73) | 28 (52) | 36 (67) | 32 (59) | 33 (61) | 25 (46) | 32 (59) |
|     ≥61 | 11 (52) | 4 (36) | 7 (64) | 6 (55) | 8 (73) | 6 (55) | 5 (45) |
| Experience (years) | | | | | | | |
|     ≤5 | 24 (49) | 13 (54) | 12 (50) | 20 (83) | 17 (71) | 14 (58) | 14 (58) |
|     6–15 | 70 (69) | 28 (40) | 44 (63) | 37 (53) | 43 (61) | 37 (53) | 45 (64) |
|     16–25 | 64 (70) | 32 (50) | 38 (59) | 39 (61) | 41 (64) | 41 (64) | 40 (63) |
|     ≥26 | 33 (63) | 17 (52) | 22 (67) | 18 (55) | 22 (67) | 19 (58) | 17 (52) |
| Use Twitter? | | | | | | | |
|     Yes | 84 (68) | 37 (44) | 44 (52) | 56 (67) | 60 (71) | 51 (61) | 53 (63) |
|     No | 108 (62) | 54 (50) | 72 (67) | 59 (55) | 64 (59) | 61 (56) | 63 (58) |
| Use Facebook? | | | | | | | |
|     Yes | 107 (64) | 52 (49) | 63 (59) | 72 (67) | 73 (68) | 67 (63) | 67 (63) |
|     No | 85 (66) | 39 (46) | 53 (62) | 43 (51) | 51 (60) | 45 (53) | 49 (58) |
| Use LinkedIn? | | | | | | | |
|     Yes | 83 (70) | 38 (46) | 47 (57) | 56 (67) | 57 (69) | 49 (59) | 53 (64) |
|     No | 109 (61) | 53 (49) | 69 (63) | 59 (54) | 67 (61) | 63 (58) | 63 (58) |
| Have personal blog? | | | | | | | |
|     Yes | 33 (75) | 17 (52) | 21 (64) | 22 (67) | 21 (64) | 22 (67) | 24 (73) |
|     No | 159 (63) | 74 (47) | 95 (60) | 93 (58) | 103 (65) | 90 (57) | 92 (58) |
| Use AE? | | | | | | | |
|     Yes | 93 (70) | 44 (47) | 55 (59) | 65 (70) | 62 (67) | 53 (57) | 57 (61) |
|     No | 99 (61) | 47 (47) | 61 (62) | 50 (51) | 62 (63) | 59 (60) | 59 (60) |
| Use RG? | | | | | | | |
|     Yes | 133 (73) | 67 (50) | 83 (62) | 84 (63) | 95 (71) | 82 (62) | 79 (59) |
|     No | 59 (52) | 24 (41) | 33 (56) | 31 (53) | 29 (49) | 30 (51) | 37 (63) |
| Use neither RG nor AE | 35 (47) | 16 (46) | 21 (60) | 16 (46) | 16 (46) | 20 (57) | 22 (63) |
| Use Mendeley? | | | | | | | |
|     Yes | 48 (75) | 21 (44) | 32 (67) | 31 (65) | 33 (69) | 29 (60) | 30 (63) |
|     No | 144 (62) | 70 (49) | 84 (58) | 84 (58) | 91 (63) | 83 (58) | 86 (60) |
| Pos. attitude to Facebook | 40 (75) | 22 (55) | 24 (60) | 25 (63) | 31 (78) | 23 (58) | 24 (60) |
| Neg. attitude to Facebook | 152 (63) | 69 (45) | 92 (61) | 90 (59) | 93 (61) | 89 (59) | 92 (61) |

| Pos. attitude to Twitter | 41 (64) | 22 (54) | 26 (63) | 23 (56) | 26 (63) | 26 (63) | 21 (51) |
|---|---|---|---|---|---|---|---|
| Neg. attitude to Twitter | 151 (65) | 69 (46) | 90 (60) | 92 (61) | 98 (65) | 86 (57) | 95 (63) |
| Pos. attitude to RG & AE | 100 (43) | 55 (55) | 61 (61) | 60 (60) | 74 (74) | 63 (63) | 56 (56) |
| Neg. attitude to RG & AE | 92 (67) | 36 (39) | 55 (60) | 55 (60) | 50 (54) | 49 (53) | 60 (65) |
| Standard deviation | 40 (14) | 19 (10) | 24 (12) | 25 (12) | 27 (13) | 23 (11) | 24 (12) |

The perception that reviewers are apt to be reluctant to criticize peers openly was highest among older professors and more veteran professors (column [3]). Early-career academics expressed the highest level of agreement with the assertion that open peer review can improve the quality of research findings (83 per cent for those with less than five years of career experience; column [4]). Survey participants who used social media agreed with the statement to varying lesser degrees that were nevertheless consistently above the study average.

A significant percentage of the survey subjects who viewed social networking in a positive light recognized problems typically associated with open peer review: 74 per cent of those who held RG and AE in high regard, and 78 per cent of those who expressed a positive attitude towards Facebook, also recognized that 'such a system would be difficult to implement in my country given its deeply rooted cultural resistance to criticism.' There was a difference of opinion between female and male study participants regarding the possibility that authors might attempt to manipulate the process by recruiting colleagues they felt would offer positive evaluations (52 per cent [males] versus 67 per cent [females]; column [6]).

*Altmetrics*

In their responses on the final section of the survey, respondents did not hold a particularly high opinion of altmetrics. Although the majority viewed them as a complement to the well-established impact factor, respondents were aware that they could be easily manipulated and that they were not sufficiently rigorous.

In Table 4, a total of 120 (41 per cent) respondents agreed with statement 'I know what altmetrics are' (see column [1]). Respondents who agreed with the other four statements concerning altmetrics—columns [2] to [5]—represent a subset of the respondents summed in column [1]. As with Tables 2 and 3, the percentages in columns [2] to [5] were calculated with the number in column [1] as the denominator. This is true for the values in all rows of Table 4.

Table 4. Respondent agreement with statements concerning altmetrics

| [1] I know what altmetrics are. [2] Altmetrics can be easily manipulated. [3] Altmetrics provide a rigorous measurement of an article's impact. [4] Altmetrics let authors distinguish their work's impact on academic audiences vs. the general public. [5] Altmetrics constitute a complement to the traditional impact factor measures. | | | | | |
|---|---|---|---|---|---|
| Participant Variable | [1] no. (%) | [2] no. (%) | [3] no. (%) | [4] no. (%) | [5] no. (%) |
| All | 120 (41) | 86 (72) | 56 (47) | 75 (63) | 94 (78) |
| Social Sciences | 74 (40) | 49 (66) | 35 (47) | 48 (65) | 58 (78) |
| Gender | | | | | |
|    Male | 67 (41) | 48 (72) | 29 (43) | 39 (58) | 50 (75) |
|    Female | 51 (40) | 36 (71) | 25 (49) | 34 (67) | 42 (82) |
| Age (years) | | | | | |
|    ≤30 | 9 (30) | 6 (67) | 5 (56) | 5 (56) | 8 (89) |
|    31–40 | 35 (49) | 24 (69) | 13 (37) | 17 (49) | 24 (69) |
|    41–50 | 43 (43) | 34 (79) | 23 (53) | 28 (65) | 37 (86) |
|    51–60 | 27 (36) | 19 (70) | 13 (48) | 20 (74) | 20 (74) |
|    ≥61 | 6 (29) | 3 (50) | 2 (33) | 5 (83) | 5 (83) |
| Experience (years) | | | | | |
|    ≤5 | 13 (27) | 11 (84) | 6 (46) | 10 (77) | 12 (92) |
|    6–15 | 36 (36) | 28 (78) | 13 (36) | 21 (58) | 25 (69) |
|    16–25 | 47 (51) | 32 (68) | 24 (51) | 29 (62) | 39 (83) |
|    ≥26 | 24 (46) | 15 (63) | 13 (54) | 15 (63) | 18 (75) |
| Use Twitter? | | | | | |
|    Yes | 58 (47) | 44 (76) | 28 (48) | 41 (71) | 50 (86) |
|    No | 62 (36) | 42 (68) | 28 (45) | 34 (55) | 44 (71) |
| Use Facebook? | | | | | |
|    Yes | 69 (41) | 52 (75) | 32 (46) | 47 (68) | 55 (80) |
|    No | 51 (40) | 34 (67) | 24 (47) | 28 (55) | 39 (76) |
| Use LinkedIn? | | | | | |
|    Yes | 51 (43) | 36 (71) | 30 (59) | 35 (69) | 44 (86) |
|    No | 69 (39) | 50 (72) | 26 (38) | 40 (58) | 50 (72) |
| Have personal blog? | | | | | |
| Yes | 19 (43) | 15 (79) | 14 (74) | 17 (89) | 18 (95) |
| No | 101 (40) | 71 (70) | 42 (42) | 58 (57) | 76 (75) |
| Use AE? | | | | | |
| Yes | 63 (47) | 47 (75) | 33 (52) | 46 (73) | 50 (79) |
| No | 57 (35) | 39 (68) | 23 (40) | 29 (51) | 44 (77) |
| Use RG? | | | | | |
| Yes | 89 (49) | 68 (76) | 46 (52) | 60 (67) | 74 (83) |
| No | 31 (27) | 18 (58) | 10 (32) | 15 (48) | 20 (65) |
| Use neither RG nor AE | 20 (27) | 12 (60) | 7 (35) | 10 (50) | 14 (70) |
| Use Mendeley | | | | | |
| Yes | 40 (63) | 30 (75) | 23 (58) | 28 (70) | 36 (90) |
| No | 80 (34) | 56 (70) | 33 (41) | 47 (59) | 58 (73) |
| Pos. attitude to Facebook | 24 (45) | 17 (71) | 9 (38) | 11 (46) | 17 (71) |
| Neg. attitude to Facebook | 96 (40) | 21 (22) | 14 (15) | 16 (17) | 18 (19) |
| Pos. attitude to Twitter | 27 (42) | 22 (81) | 16 (59) | 18 (67) | 22 (81) |
| Neg. attitude to Twitter | 93 (40) | 16 (17) | 7 (8) | 9 (10) | 13 (14) |
| Pos. attitude to RG & AE | 68 (49) | 25 (37) | 20 (29) | 22 (32) | 25 (37) |
| Neg. attitude to RG & AE | 52 (33) | 13 (25) | 3 (6) | 5 (10) | 10 (19) |
| Standard deviation | 28 (7) | 20 (27) | 13 (21) | 17 (23) | 21 (29) |

An analysis of survey responses by age group revealed that 49 per cent of the participants aged 31–40 and 36 per cent of the study population aged 51–60 were familiar with altmetrics. Nevertheless, a comparison of responses on this topic from the angle of professional experience indicates that early-career researchers know less about this system of measurement than their more veteran counterparts. A breakdown of the study population by years of professional experience indicates that only 27 per cent of respondents with less than five years of experience in their field were aware of how altmetrics function compared with 36 per cent with 6–15 years, 51 per cent with 16–25 years, and 46 per cent with more than 26 years of professional practice.

**Discussion**

Our findings, which indicate that most contributors to Spanish scholarly journals hold a favourable opinion of OA, generally coincide with those of similar studies.[30] The individuals whom we surveyed appear to agree with the editors of library and information science journals who, in one study, fully expected OA to become the dominant model in scholarly publishing within the next five to ten years.[31] Our respondents were nevertheless less aware and more cautious about two other emerging trends in scholarly publishing not as well tested—open peer review and altmetrics.

Although greater percentages of older respondents and respondents with more years of experience recognized the advantages of OA, above average percentages of the same groups also agreed with its drawbacks. This emphasis on the potential negative aspects of OA suggests a certain conservatism on their part that may be related to their prolonged acquaintance with other publishing frameworks and practices, a circumstance that could explain their lower level of confidence in the concept. Nevertheless, a link was detected between the length of respondents' professional experience and their awareness and understanding of altmetrics.

A positive attitude toward social networks did not necessarily translate into enthusiasm for emerging trends in scholarly publishing. Respondents claiming to have a positive opinion of Twitter demonstrated the lowest level of support of any segment of the study population for the assertion that OA enhances the visibility of scholarly articles. One possible explanation for this seeming contradiction could be that these authors attributed the visibility of their scholarly output as much to their personal efforts to promote it as to the effect of a journal's OA policy.

Study participants with positive attitudes toward social media tended to agree with the statement focusing on the negative aspects of open peer review, especially the possibility of evaluations being tainted by personal biases and agendas. Respondents with negative feelings about RG and AE supported this notion to a lesser degree (39 per cent) than those who viewed these sites in a positive light, 55 per cent of whom agreed with this assertion.

Our findings concerning the attitudes of academics at an early stage of their career fell in line with those of other studies that have noted the widespread reluctance of younger scholars to accept open peer review practices.[32] Previous studies analyzing the attitudes of editors at a Spanish scholarly journal[33] and another conducted in Turkey[34] found widespread resistance to open peer review.

One point on which we detected a difference of opinion among social media users was the question of whether or not open peer review actually improves the quality of research findings. Our findings indicate a link between familiarity with these platforms and more positive attitudes on the part of academics regarding open review.

Regarding the doubts and drawbacks about open peer review, it should be kept in mind that few academic journals devoted to the social sciences and none of the Spanish publications analyzed for this study have adopted those practices, so the context in which the participants responded to statements about it was purely speculative. The fact that respondents who engage in academic social networking are aware of the positive aspects of open peer review in no way precludes the possibility of their being equally aware of its downsides, one of which is its potential vulnerability to manipulation.

Female respondents expressed more doubts about the open review process than their male colleagues. This could be a result of the various forms of gender discrimination they suffer,[35] which range from their articles being cited less often than those written by male authors,[36] to a lower probability that their requests for funding will receive positive evaluations,[37] and to a lesser frequency with which they are invited to be keynote speakers at conferences related to their fields.[38] Female study participants' relative lack of enthusiasm about open review, a process that does not protect authors' anonymity, could be related to a perception that traditional blind peer review constitutes a hedge against gender bias.

The general lack of knowledge about altmetrics detected among participants in this study is line with previous research focusing on other types of university professionals such as librarians.[39] Almost half of the RG users in the survey population (49 per cent) claimed to be familiar with the concept. As RG has developed the ResearchGate Score, a proprietary system for measuring scientific reputation[40] based on a mix of altmetrics and more traditional bibliometrics,[41] it is not particularly surprising that scholars using that platform would have a better than average grasp of what altmetrics are about.

The findings of this study fall in line with those of others indicating that the open peer review process tends to be more frequently employed in the biomedical sciences than in the social sciences.[42] (Note that a small proportion of our participants represented the sciences, while the majority represented the social sciences.) Traditionally, experimental sciences have been in the vanguard of changes in scholarly communication, as it happened with OA. That could explain that researchers in the social sciences and humanities are more reluctant to these new features in the scholarly publishing process as the open peer review. It would nevertheless be worth exploring whether the implementation of open peer review practices in classroom settings[43] could change future researchers' attitudes toward the practice.

**Limitations and Further Research**

The extent to which the findings of this study can be generalized is admittedly limited. It must be kept in mind that the survey universe used in our research was restricted exclusively to authors of articles published in Spanish scholarly journals. As this choice of study design predetermined that the majority of subjects participating lived and worked in Spain, what this article offers is a snapshot of opinions and attitudes of academic authors in essentially one country. It is nevertheless worth noting that the scope of this study was not restricted to a single institution or area of knowledge. The data generated can therefore be used to draw comparisons between the status quo on this topic in other countries and cultural contexts. As our response rate from academics working in the social sciences was higher than that of academics in other disciplines, the opinion of these other groups are underrepresented in our findings.

We suggest that further studies measure academics' specific use of each social media service with an eye to determining whether the strategies they employ lead to differing perceptions of open access, open peer review, altmetrics and other changes in the scholarly publishing landscape.

**Acknowledgements**

This work has been funded by Universidad Internacional de La Rioja (grant no. B0036-1718-609).

Authors would like to thank Rafael Repiso for his views on the initial conception of the study, Ana Belén Calvo for her suggestions on the first draft of the paper, the anonymous reviewers for their suggestions on the initial manuscript and specially to the authors who generously contributed to the survey.

---

[1] Xin Gu and Karen Blackmore, 'Characterisation of Academic Journals in the Digital Age,' *Scientometrics* 110, no. 3 (2017): 1333–50.

[2] Chris R. Triggle and David J. Triggle, 'From Gutenberg to Open Science: An Unfulfilled Odyssey,' *Drug Development Research* 78, no. 1 (2017): 3–23.

[3] Roger Clarke, 'The Cost Profiles of Alternative Approaches to Journal Publishing,' *First Monday* 12, no. 12 (2007), doi:10.5210/fm.v12i12.2048.